\documentclass[5p,times]{elsarticle}

\usepackage[utf8]{inputenc}
\usepackage[T1]{fontenc}
\usepackage{amsmath,amssymb,amsfonts,mathrsfs}
\usepackage{algorithmic}
\usepackage{graphicx}
\usepackage{textcomp}
\usepackage[table,dvipsnames]{xcolor}
\usepackage{booktabs} %
\usepackage{rotating}
\usepackage{subcaption}
\usepackage{multirow}
\usepackage{listings}
\def\BibTeX{{\rm B\kern-.05em{\sc i\kern-.025em b}\kern-.08em
    T\kern-.1667em\lower.7ex\hbox{E}\kern-.125emX}}

\usepackage{xspace}
\newcommand{\ie}{i.e.,\xspace}
\newcommand{\eg}{e.g.,\xspace}
\newcommand{\etc}{etc.\xspace}

\newcommand{\mn}{MareNostrum~5\xspace}

\newcommand{\fall}{FALL3D\xspace}
\newcommand{\mfall}{MiniFALL3D\xspace}

\xspaceaddexceptions{=}

\usepackage{listings} %
\definecolor{col1}{HTML}{1E88E5}
\definecolor{col2}{HTML}{D81B60}
\definecolor{col3}{HTML}{43A047}
\definecolor{col4}{HTML}{F4511E}
\definecolor{col5}{HTML}{205B74}

\usepackage{pifont}

\usepackage[colorlinks=true, allcolors=blue, bookmarks=false]{hyperref}

\RequirePackage[normalem]{ulem}
\RequirePackage{color}\definecolor{RED}{rgb}{1,0,0}\definecolor{BLUE}{rgb}{0,0,1}

\providecommand{\DIFdel}[1]{{\protect\color{RED}\sout{#1}}}
\providecommand{\DIFdel}[1]{}

\journal{PPAM 2024 - Advances for HPC Systems}

\begin{document}

\begin{frontmatter}

\title{Exploring RISC-V Long Vector Capabilities: A Case Study in Earth Sciences}

\author[bscmainaddress]{Fabio Banchelli\corref{mycorrespondingauthor}}
\cortext[mycorrespondingauthor]{Corresponding author}
\ead{fabio.banchelli@bsc.es}

\author[bscmainaddress]{David Jurado}
\author[bscmainaddress]{Marta Garcia-Gasulla}
\author[bscmainaddress]{Filippo Mantovani}

\address[bscmainaddress]{Barcelona Supercomputing Center, Plaza Eusebi Guell, 1-3, 08034 Barcelona (Spain)}

\begin{abstract}

This paper investigates the performance of Earth Sciences codes, specifically SeisSol and MiniFALL3D, on a RISC-V-based CPU featuring a long vector processing unit.
The study focuses on optimizing these applications for improved computational efficiency while maintaining portability.
For SeisSol, we explore batched GEMM implementations to enhance performance by leveraging instruction-level parallelism.
MiniFALL3D's optimization involves improving vectorization by modifying the source code, such as replacing functions with subroutines and flattening multidimensional arrays.
The vectorization process is always left to the compiler to ensure code portability.
The study is conducted using both a software emulator and a hardware prototype of the RISC-V vector architecture called EPAC.
The performance of both applications is evaluated across different HPC platforms, including EPAC (based on RISC-V), MareNostrum 4 (powered by Sapphire Rapids CPUs), and the NEC SX-Aurora Tsubasa accelerator.
We aim to provide insights into adapting Earth Sciences codes for modern high-performance computing systems while demonstrating the potential of RISC-V vector architectures.
Ultimately, all modifications made to improve performance on the RISC-V long vector architecture are shown to be beneficial on other HPC architectures with different vector capabilities.
This highlights the importance of maintaining code portability while relying on the compiler's powerful auto-vectorization capabilities.

\end{abstract}

\begin{keyword}
RISC-V
\sep
Long Vector Architectures
\sep
High-Performance Computing
\sep
Batched GEMM
\sep
Seissol
\sep
\mfall
\end{keyword}

\end{frontmatter}

\section{Introduction}\label{secIntro}

The RISC-V open-source instruction set architecture (ISA) has emerged as a compelling alternative to proprietary architectures in high-performance computing (HPC). Its flexibility, scalability, and transparency make it particularly attractive for scientific and research applications. A key feature of RISC-V that enhances its suitability for HPC workloads is its support for vector computation through the RISC-V Vector Extension (RVV). This extension introduces scalable vector instruction capabilities, enabling efficient exploitation of data-level parallelism in scientific applications.
The RISC-V Vector Extension provides a framework for executing vector operations, which are critical for accelerating data-parallel workloads. The first version, RVV 0.7.1, was introduced in June 2019, followed by the ratification of RVV 1.0 in September 2021, which is now the standard. RVV 1.0 refines the earlier draft by standardizing essential features such as register grouping, mask registers, and support for variable-length vectors. Unlike fixed-length vector architectures such as x86's SSE, AVX2, and AVX512 (which support 128-bit, 256-bit, and 512-bit vectors, respectively), RVV adopts a vector length agnostic programming model similar to the one adopted by Arm SVE. This model abstracts hardware-specific vector lengths, allowing programmers to write architecture-agnostic vector code. The vector length determines the number of elements processed per instruction, with higher vector length values improving throughput for data-parallel workloads.

The European Processor Initiative\footnote{\url{https://www.european-processor-initiative.eu/}} (EPI) has embraced RISC-V's vector architecture to develop the EPI Accelerator (EPAC), a high-performance computing accelerator able to operate as a standalone compute node. EPAC integrates a scalar CPU with a Vector Processing Unit (VPU) that supports RVV 0.7.1 (recently updated to RVV 1.0). The VPU features 32 vector registers up to 16 kbit wide, enabling each vector instruction to operate on up to 256 double-precision elements. This capability represents an extreme design point compared to other mainstream architectures available in HPC and is particularly advantageous for HPC applications requiring high computational throughput.
The hardware design developed within the EPI project includes a RISC-V micro-tile composed of an Avispado scalar core (developed by Semidynamics\footnote{Semidynamics. \url{https://semidynamics.com/}})
connected to a Vitruvius VPU~\cite{minervini2023vitruvius} with eight lanes. Each lane incorporates a Floating Point Unit (FPU) developed by the University of Zagreb~\cite{kovavc2023faust}. The micro-tile also includes a Home Node and an L2 cache, designed by Chalmers\footnote{Chalmers University of Technology. \url{https://www.chalmers.se}} and FORTH\footnote{FORTH Institute of Computer Sciences. \url{https://www.ics.forth.gr/carv}}, respectively.

Preparing scientific codes for emerging architectures is a complex task, as demonstrated by the work conducted within the ChEESE Centre of Excellence (CoE) for Exascale Supercomputing in Solid Earth. Coordinated by Spain’s CSIC, ChEESE integrates 16 European institutions to optimize computational workflows for seismic, volcanic, and tsunami hazard modeling. Key scientific codes employed by ChEESE include:
SeisSol (LMU Munich),
SPECFEM3D (CNRS),
Tandem (LMU Munich),
xSHELLS (CNRS),
HySEA (University of Malaga),
FALL3D (CSIC),
OpenPDAC (INGV),
LaMEM (University of Mainz),
pTatin3D (Sorbonne Université),
ELMER/ICE (CSC – IT Center for Science Ltd).
These codes leverage adaptive mesh refinement, GPU acceleration, and parallel scalability to enable high-resolution simulations of coupled Earth systems. ChEESE enhances urgent computing capabilities for real-time hazard mitigation and integrates workflows with EuroHPC infrastructure, emphasizing performance-portable frameworks and training in exascale computational geoscience. In this paper, we focus on studying two of these codes, SeisSol and MiniFALL3D, when running on the EPAC architecture.

This paper is an extension of~\cite{banchelli2025batched} presented at PPAM~2024. Compared to the original submission, we updated the original data comparing with \mn and we extended the work including a study of an additional earth science HPC application.
This paper makes the following contributions:
{\em i)} Batched GEMM Optimization for RISC-V: We present a portable, RVV-friendly implementation of batched GEMM (General Matrix Multiply) that achieves up to $32.6\times$ speedup over OpenBLAS on EPAC. This optimization leverages compile-time matrix sizes and strided memory accesses.
{\em ii)} Vectorization of MiniFALL3D: We detail code modifications that increase the Vector Mix from 0.36\% to 7.94\% and Vector Activity from 9.05\% to 72.4\%, resulting in a $6.32\times$ speedup on EPAC.
{\em iii)} Cross-Architecture Analysis: We evaluate the performance of our optimizations on Intel Sapphire Rapids (AVX-512) and NEC SX-Aurora, demonstrating their portability across architectures.

The remainder of this paper is organized as follows:
Section~\ref{secMethodology} introduces the methodology, software, and hardware tools used for the vectorization study.
Section~\ref{secSeissol} details SeisSol’s GEMM optimizations.
Section~\ref{secMiniFall3D} explores MiniFALL3D’s vectorization challenges and solutions.
Section~\ref{secRelatedWork} presents the related work.
Section~\ref{secConclusions} concludes with insights on vectorization, portability considerations, and the future role of RISC-V in HPC systems.

\section{Methodology}\label{secMethodology}

\subsection{Hardware platform}
All development and experiments are performed in the fpga-sdv cluster~\cite{sdv-paper}.
This system is based on an FPGA design that implements EPAC, a RISC-V scalar core tightly coupled with a vector unit that support up to~$256$ double-precision elements per instruction and can process up to~$8$ elements per cycle.
The ISA extensions available in this system are~\texttt{rv64gcv}, with the vector extension being \texttt{rvv0.7}.
The core runs at a frequency of~$50$~MHz and runs a standard Ubuntu~$22.4$ Linux image.

\subsection{Software environment}
We use a Clang-based compiler developed within the EPI project that is able to auto-vectorize C, C++, and Fortran codes targeting the RISC-V vector extension.
This modified compiler supports both \texttt{rvv0.7} and \texttt{rvv1.0}.
This auto-vectorization can be achieved without any code modifications.
There are also some compiler hints in the form of pragmas which improve the auto-vectorization that are compatible with the upstream version of Clang (\eg \texttt{loop vectorize}).
Finally, there is also a set of compiler intrinsics to manually vectorize the code which are specific to this version of Clang.
In this work, we strongly advocate for the use of compiler hints and to stray away from platform-specific code in favor of maintaining portability.

In addition to the compiler, the scientific codes studied in this work have other software dependencies.
Common scientific libraries are available for EPAC.
For the case of SeisSol, the most relevant dependency is a BLAS library.
Section~\ref{secSeissol} lists the available alternative, their limitations, and their impact in performance.
For \mfall, NetCDF is required for compiling, but it is not relevant for the performance evaluation.
Section~\ref{secMiniFall3D} details how NetCDF is disabled for our experiments.

\subsection{Tracing tools}

Throughout this work, we use timelines as a way to visualize relevant aspects of code executions.
These timelines are a product of a set of tracing tools integrated into our development infrastructure.
Traces are then translated again to be visualized using Paraver~\cite{pillet1995paraver}, a trace visualizer developed at the Barcelona Supercomputing Center.

\paragraph{RAVE traces}
We use QEMU to emulate a RISC-V system with a vector unit.
RAVE is a QEMU plugin that monitors the emulated executions and produces traces that can be later visualized and studied~\cite{vizcaino2024rave}.
RAVE traces are not cycle accurate, since they originate from an emulated execution, but contain information such as the type of vector instructions executed and the code regions that contain them.
In this work, all timelines that show code phases or types of instructions have been generated with RAVE.

\paragraph{Signal traces}
We leverage a custom hardware tracer embedded within the FPGA design that can spy the value of cherry-picked signals at each clock cycle.
The tracer is configured before the application execution and triggers automatically when a given instruction is executed (\ie \texttt{vor.v}).
It is also completely decoupled from the core, meaning that it has no impact on the performance of the application and they are cycle accurate.
The traced data is read from the FPGA after the application has ended and transformed into a human-readable format.
In this work, all timelines that show hardware resources such as the arithmetic unit or in-flight memory operations have been generated with the hardware tracer.

\subsection{Vectorization metrics}

In this work we base our porting efforts on the evaluation methodlogy proposed in~\cite{short-reasons}.
The most relevant metrics for our study are detailed here.

\paragraph{Average Vector Length (AVL)}
Given a variable vector length system, the AVL is the vector length of each vector instructions averaged throughout a certain region under study.
Since our target platform has a maximum vector length of~$256$ double-precision elements, the target AVL should match this figure.
Increasing the AVL is our first priority when porting codes to long-vector architectures.

\paragraph{Vector Mix}
We call the fraction of vector instructions with the respect to the total instructions the Vector Mix.
This metric is represented as a percentage where~$0\%$ means that there are no vector instructions, and $100\%$ means that all instructions are vector.
While for simple codes, such as benchmarks, it is easy to write kernels that have a Vector Mix of~$100\%$, it is unfiseable to have such a high value for real scientific applications.
A general rule of thumb is to target Vector Mix above~$20\%$.

\paragraph{Vector Activity}
Similar to Vector Mix, the Vector Activity represents the fraction of cycles spend computing vector instructions with respect to the total cycles of a program execution.
This metric is also represented as a percentage and a general rule of thumb is to target Vector Activity above~$80\%$.

Section~\ref{secSeissol} focuses on optimizing GEMM computations by increasing the AVL,
while Section~\ref{secMiniFall3D} studies how to improve the Vector Mix and Vector Activity metrics.

\section{SeisSol}\label{secSeissol}

SeisSol~\cite{seissol} is
a software package for simulating wave propagation and dynamic rupture based on the arbitrary high-order accurate derivative discontinuous Galerkin method (ADER-DG).
The official repository of SeisSol~\cite{seissol-repo} includes multiple build targets, but in this work we focus on \texttt{SeisSol-proxy}.
This target compiles a subset of the application which is representative of real workloads but simpler to study and analyze.
When building the SeisSol-proxy application,
even if running with a single core, the developers advise to enable MPI and OpenMP.
All optional software dependencies are disabled.
The \texttt{ORDER} parameter changes the mathematical problem that is solved by the application.
In this work, we focus on a single value for this parameter, \texttt{ORDER=4}, which was provided by the developers.

\subsection{Execution structure}
\begin{figure}[!htbp]
  \centering
  \includegraphics[width=.9\columnwidth]{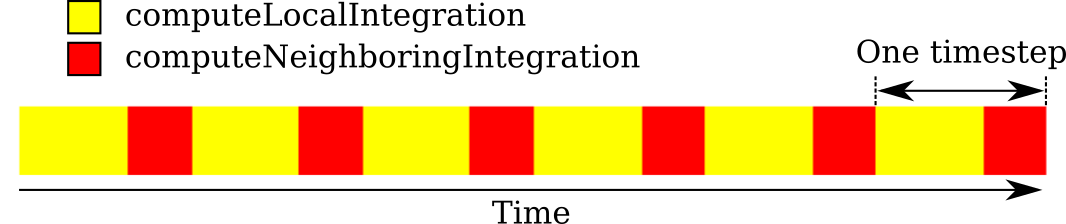}
  \caption{Timeline of six timesteps in SeisSol. Yellow regions corresponds to {\em computeLocalIntegration} while red regions correspond to {\em computeNeighboringIntegration}.}
  \label{figSeissolTimestepStructure}
\end{figure}

The execution flow of the SeisSol-proxy application is divided into timesteps.
The number of timesteps is a runtime parameter.
Each timestep consists in two distinct phases: {\em computeLocalIntegration} and {\em computeNeighboringIntegration}.
Figure~\ref{figSeissolTimestepStructure} shows an execution timeline of SeisSol.
The $x$-axis represents time, and colored regions correspond to different execution phases.
We observe that the {\em computeLocalIntegration} region, shown as yellow, is the dominant phase throughout the execution.
In this work, we focus our efforts into studying and optimizing only the {\em computeLocalIntegration}.

Diving into the structure of the {\em computeLocalIntegration} region ({\em local-integration} from hereon), we find that the program iterates through a number of cells, which is a runtime parameter.
For each cell, two functions are executed: \texttt{computeAder} and \texttt{computeIntegral}.
One level deeper, we find functions preceded by the namespace \texttt{kernel::}, which correspond to auto-generated code which is determined during the build process of SeisSol.
If the user has selected an implementation optimized for a given architecture or selected a specific GEMM (General Matrix-Matrix multiplication) library, the build system will generate different kernels.
We further categorize the kernels into two types: GEMM-based kernels and the rest.
Table~\ref{tabSeissolKernelCategories} summarizes the contribution of each kernel category to the execution time.

\begin{table}[htbp]
\centering
\caption{Categorization of kernels in SeisSol-proxy.} \label{tabSeissolKernelCategories}
  \centering
  \begin{tabular}{lr}
    \textbf{Region}                                & \textbf{\%Cycles}\\ \midrule
    \texttt{Timestep}                              &             - \\
    \texttt{..computeLocalIntegration}             &       $66.63$ \\
    \texttt{....kernel::derivativeTaylorExpansion} &       $22.24$ \\
    \texttt{....GEMM-based kernels}                &       $35.67$ \\
    \texttt{....Other}                             &       $ 7.84$ \\
    \texttt{..computeNeighboringIntegration}       &       $33.37$ \\
  \end{tabular}%
\end{table}

As shown in Table~\ref{tabSeissolKernelCategories}, the GEMM-based kernels are the most time consuming regions of code, representing~$35.67\%$ of the execution time in one timestep.
In the following sections, we focus on these type of kernels.

\subsection{GEMM-based kernels}\label{secDgemm}

\paragraph{General structure}
All GEMM-based kernels invoke double-precision GEMM calls (\eg \texttt{cblas\_dgemm} for OpenBLAS).
The number of GEMMs and sizes of the matrices vary from kernel to kernel, but are known at compile time.
Furthermore, some kernels allocate temporary buffers in the stack.
These temporary buffers are of a fixed size known at compile time and will be relevant later in this work.
The specific calls depend on the GEMM tools selected during the build process.

OpenBLAS~\cite{openblas-repo} is a widely used open-source BLAS implementation that provides architecture-specific optimizations.
Support for the rvv0.7 extension is limited since the main development focus is towards rvv1.0.
The version available in our environment is~0.3.20 and it is not vectorized for rvv0.7.

BLIS~\cite{BLIS1,BLIS2} is a portable software framework for instantiating high-performance BLAS-like dense linear algebra libraries.
BLIS uses a different API as other BLAS libraries although it also includes a BLAS compatibility layer to increase code portability.
There is no official rvv0.7 implementation of BLIS.
The version available in our environment is~0.8.1 and it is not vectorized.

Eigen~\cite{eigen-repo} is a C++ template library for linear algebra.
Since Eigen is a header-only library, it is compiled together with the application.
This means that by enabling auto-vectorization for SeisSol, we are also enabling it for Eigen.
It also means that there is no way to only enable auto-vectorization for Eigen code without changing the build system of SeisSol.

\subsection{Performance out-of-the-box}
In this section we present the performance {\em GFLOPs (hardware)} reported by the application.
This metric measures the amount of floating-point operations per second performed during an execution.
Unless stated otherwise, all runs were performed setting the number of cells to~$10000$.

\begin{figure}[!htbp]
    \centering
    \includegraphics[width=.9\columnwidth]{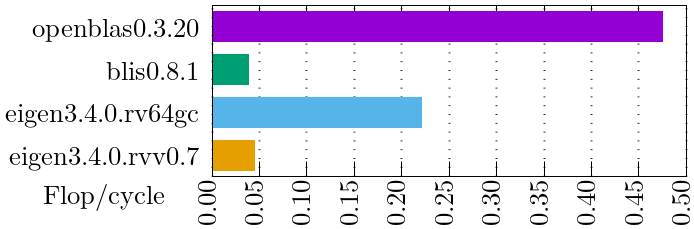}
    \caption{Performance of SeisSol in EPAC using different GEMM libraries.}
    \label{figSeissolGflops}
\end{figure}

Figure~\ref{figSeissolGflops} show the performance of SeisSol running in EPAC at~$50$~MHz.
Each bar represents the average performance across five runs with a standard deviation of less than~$3\%$.
We observe that the performance of SeisSol greatly varies depending on the GEMM library in use.
From worst to best: BLIS, Eigen, and OpenBLAS.
Surprisingly, we observe that the performance of the auto-vectorized build using the Eigen library yields a much lower performance compared to the scalar build.
Furthermore, the auto-vectorized build has a performance under~$0.05$~Flop/cycle, which is much lower of the theoretical peak performance the vector unit~($16$~Flop/cycle).

Further investigation shows that the compiler emits vector instructions when compiling Eigen code, but the C++ templated nature and high level of abstraction of the code makes it very difficult to actually leverage long vectors.
The two most relevant types of instructions that are generated are \texttt{vfredsum} and \texttt{vfmadd}.
The first one is a reduction operation and operates on vectors of maximum size ($VL=256$).
For each \texttt{vfredsum}, the compiler must prepare a register with indices using the \texttt{vid} instruction, convert the datatype of the indices with \texttt{vwcvtu} and change the Vector Lenght with \texttt{vsetvli}.
Thus, \texttt{vfredsum} is a very costly operation.

\begin{figure}[!htbp]
  \centering
  \includegraphics[width=.9\columnwidth]{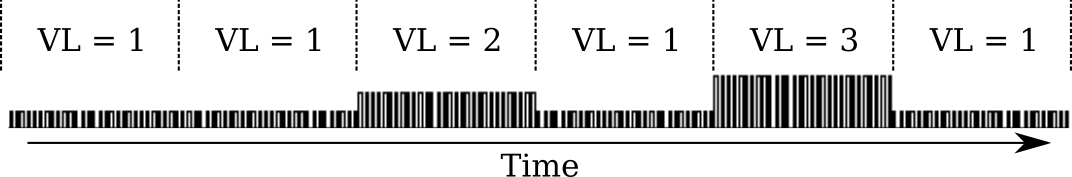}
  \caption{VL of \texttt{vfmadd} instructions during the kernel \texttt{derivative::execute1}.}
  \label{figSeissolVfmadVl}
\end{figure}

Regarding \texttt{vfmadd}, we know that this instruction is the main bulk of the matrix-matrix useful computation.
Leveraging a wide vector unit with \texttt{vfmadd} is crucial for maximizing performance of the matrix-matrix multiplication.
Figure~\ref{figSeissolVfmadVl} shows the Vector Length (VL), in double-precision elements, of \texttt{vfmadd} throughout \texttt{kernel::derivative::execute1}.

We observe six regions of the same duration that correspond to the six GEMM operations inside the kernel.
In here lies the main performance limiter: the VL of \texttt{vfmadd} instructions is between one and three double-precision elements, when the vector unit can support up to~256.
The cost of issuing one instruction to the vector unit is very high compared to a scalar instruction, but this cost is usually hidden by operating on multiple elements per instruction.
Paying the issue cost for each element defeats the purpose of using the vector unit.

To better leverage hardware based on long vectors, we need to expose more instruction level parallelism.
To achieve this, the code must expose more {\em work}.
For example, merging the computation of multiple cells into a single function scope.
The following section presents an implementation of GEMM that processes batches of matrices, corresponding to multiple cells, instead of a single pair of matrices (single cell).

\subsection{Batched GEMMs}\label{secBatchedDgemm}

\paragraph{Standards and problem constraints}

There are proposed standards for batched GEMM and BLAS libraries in the literature~\cite{dongarra2016proposed,dongarra_batchedgemm,DONGARRA2017495}.
However, there seems to be no common ground or standard yet.
Most of the proposals, define a function header in which matrices cannot be assumed to be placed consecutively in memory (\eg defining the parameter \texttt{double** A} instead of \texttt{double* A}).
With this generic API, there is no room for improvement for codes which do allocate batches of matrices consecutively.
Other constraints imposed by using a generic standard include assuming that matrix sizes (\ie \texttt{N}, \texttt{M}, \texttt{K}) and scalar components (\ie \texttt{alpha} and \texttt{beta}) vary throughout the batch.
Furthermore, there is no consideration for codes in which some parameters are known at compile time.
There are also some implementations of such standards that have been tuned to specific micro-architectures~\cite{mkl_batchgemm,cublas_batched}.
However, these implementations are either closed-source not portable to other architectures.

In the case of SeisSol, matrix sizes are {\em i)} constant throughout the batch, {\em ii)} known at compile time.
Scalar components \texttt{alpha} and \texttt{beta} are constant throughout the batch.
Depending on the kernel, memory allocation of matrices \texttt{A}, \texttt{B}, and \texttt{C} throughout the batch can be one of three types:
  {\em Constant} if the same matrix is used for the whole batch (\texttt{c}).
  {\em Strided}  if matrices are contiguous in memory (\texttt{s}).
  {\em Indexed}  if matrices are not contiguous in memory (\texttt{i}).

Considering the constraints of the available batched BLAS APIs and that, to the best of our knowledge, there are no implementations compatible with the RISC-V architecture,
we propose an implementation of batched GEMM calls that
{\em i)} is written in plain C, which means that it is portable to different architectures; and
{\em ii)} leverages the optimization opportunities exposed by SeisSol (\ie constant matrix sizes, non-indexed memory layouts, \etc)
Our proposal is equally generic as any other GEMM implementation, since it can solve any matrix-matrix multiplication, but requires to know the size of the given matrices at compile time and generate the corresponding kernels.
This is a step that is already in use in SeisSol for all kinds of mathematical kernels, but may limit the usefulness of our proposal in other scientific codes.

The function header that we proposed is based on the standard \texttt{cblas\_dgemm} but making the matrix size parameters (\texttt{N}, \texttt{M}, and \texttt{K}) part of the name of the function.
We also add the type of access to each matrix as a suffix of the function name.
Thirdly, the matrix format (\ie column major and row major) are also moved to the function name.
Lastly, we add a parameter \texttt{E} with corresponds to the size of the batch.
Listing~\ref{lstBbdgemmApi} shows an example of the proposed API.

\begin{lstlisting}[caption={Proposed API for Batched GEMMs},label={lstBbdgemmApi},language=C++]
// N = 2, M = 3, K = 4
// Matrix A: 2x4, Constant
// Matrix B: 4x3, Indexed
// Matrix C: 2x3, Strided
void bbdgemm_ColMajor_2_3_4_cis(long E,
  double alpha, const double* A, long lda,
  const double* const* B, long ldb, double beta,
  double* C, long ldc);
\end{lstlisting}

The reader should note that constant and strided access types make the matrix parameter \texttt{double*} while indexed makes it \texttt{double**}.
This is because the first access type assumes matrices are stored contiguously in memory,
while the later assumes a vector of pointer to matrices.
This difference is of crucial importance, since it allows certain load and store compile time optimizations for contiguous matrices that are not possible for the vector of pointer to matrices.

\paragraph{Implementation}
A naive implementation of a Batched GEMM is to write four nested loops iterating through the size of the batch \texttt{E}, and then each matrix size \texttt{M}, \texttt{N}, \texttt{K}.
Classical optimizations include {\em i)} loop reordering to leverage spatial locality, and {\em ii)} matrix tiling to leverage fixed-sized SIMD registers.
In the case of SeisSol and a generic long vector architecture, these two optimizations are not optimal:
Since the matrix sizes of our use case are small (\ie $20\times10$ elements at most), no loop reordering of \texttt{M}, \texttt{N}, and \texttt{K} will allow for full usage of a long vector.
In addition, matrix tiling for such small matrices is not beneficial, and we do not want to limit code portability by coding for a specific vector length.

To leverage long vectors, we need to expose more instruction level parallelism to the compiler.
To do so, our implementation writes a single loop over the batch size \texttt{E} which is a runtime parameter in the order of~$10^4$.
The body of this loop is a full unroll of the three traditional \texttt{M}, \texttt{N}, \texttt{K} loops.
Listing~\ref{lstBbdgemmImplementation} shows an example implementation for a specific case.

\begin{lstlisting}[caption={Plain C implementation of Batched GEMMs},label={lstBbdgemmImplementation},language=C++]
void bbdgemm_ColMajor_2_2_2_cis(long E, /* ... */){
  long sizeC = 2 * ldc;

  #pragma clang loop vectorize(assume_safety)
  for (long e = 0; e < E; e++) {
    double vA_0_0 = A[(0*lda+0)];
    double vA_0_1 = /* Load all elements of A */
    double vB_0_0 = B[e][(0*ldb+0)];
    double vB_0_1 = /* Load all elements of B */

    double rC_0_0 = 0;
    rC_0_0 = vA_0_0 * vB_0_0 + rC_0_0;
    rC_0_0 = vA_0_1 * vB_1_0 + rC_0_0;
    rC_0_0 = rC_0_0 * alpha;
    double rC_1_0 = /* Operate all elements of A and B */

    double vC_0_0 = C[e*sizeC+(0*ldc+0)];
    rC_0_0 = vC_0_0 * beta + rC_0_0;
    C[e*sizeC + 0*ldc+0] = rC_0_0;
    double vC_1_0 = C[e*sizeC+(0*ldc+1)];
    rC_1_0 = vC_1_0 * beta + rC_1_0;
    C[e*sizeC + 0*ldc+1] = rC_1_0;
    /* Compute and store all elements of C */
  }
}
\end{lstlisting}

The pragma annotation of the loop acts as a hint for the compiler to try to vectorize the loop.
The reader should note that there is no architecture-specific code in our implementation.
If the pragma is not recognized by a compiler, it will simply be ignored and the compilation will proceed as normal.

In our case, our Clang-based compiler is able to generate vector code fully leveraging the long vectors of our system by processing~$256$ scalar iterations in a single vectorized iteration.
Matrices are accessed using the following vector instructions:
  matrix A, vector-scalar instructions (\eg \texttt{vfmacc.vf}) since it is constant throughout the batch and it only needs to be allocated in scalar registers.
  Matrix B, indexed vector loads (\texttt{vlxe.v}) since the matrices are not contiguous in memory.
  Matrix C, strided vector loads and stores (\texttt{vlse.v} and \texttt{vsse.v}) since the matrices are contiguous in memory.

\begin{figure}[!htbp]
    \centering
    \includegraphics[width=.9\columnwidth]{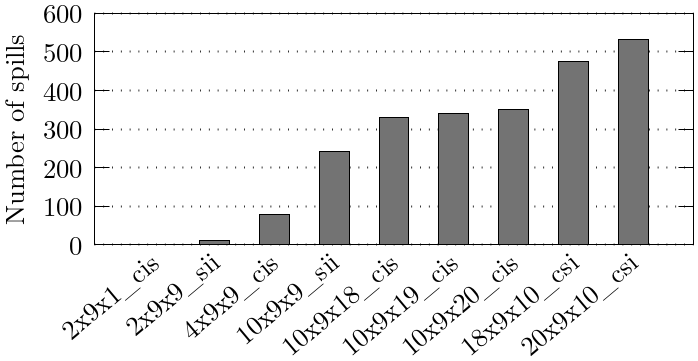}
    \caption{Number of register spills when increasing the matrix sizes.}
    \label{figSeissolRegisterSpilling}
\end{figure}

\paragraph{Register spilling}
Our implementation of the Batched GEMM targets small matrices.
The bigger the matrix size, the more hardware registers need to be allocated within one iteration of the loop.
At a certain point, there are no registers left and the compiler is forced to push some of the contents of the registers to main memory.
This effect is known as register spilling and it may have a negative impact in performance.
Figure~\ref{figSeissolRegisterSpilling} shows the number of register spills per loop iteration of our implementation when increasing the matrix size.

We observe that the number of spills shoots up with bigger matrices.
For each register spill, the compiler introduces a vector unit-strided store (save the register) and a vector unit-strided load (recover the register).
Paying the cost of these two instructions, although being memory operations, is beneficial when compared to indexed memory operations.
Thus, by fully unrolling the matrix multiplication loops, we minimize the amount of costly memory instructions (\texttt{vlxe.v} and \texttt{vlse.v}) in exchange of introducing a cheaper alternative (\texttt{vle.v}).

\begin{figure}[!htbp]
    \centering
    \includegraphics[width=\columnwidth]{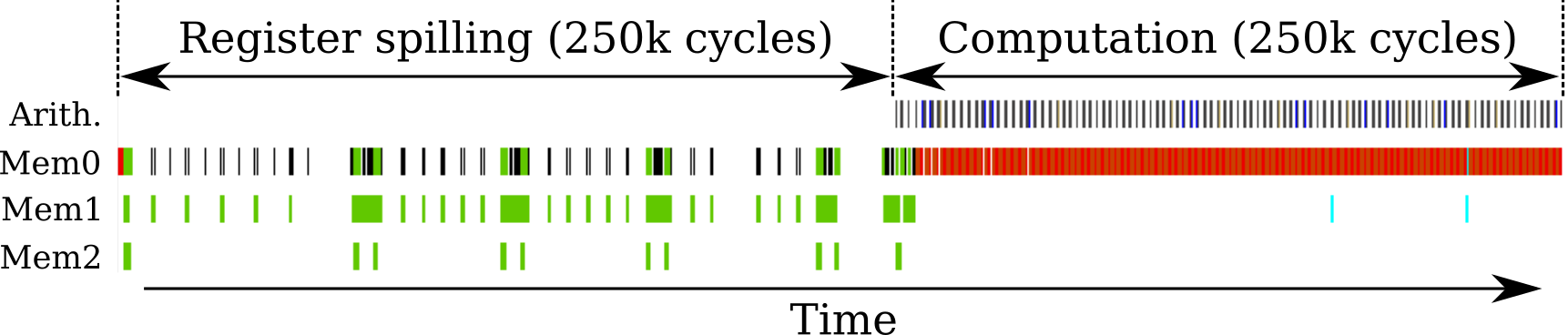}
    \caption{Instruction timeline of \texttt{20\_9\_10\_csi} (all).}
    \label{figSeissolTimelineAll}
\end{figure}

Figure~\ref{figSeissolTimelineAll} shows a timeline of the case \texttt{20\_9\_10\_csi}, which is our worst input case.
The $x$-axis represents time and each row represents a hardware resource.
The first resource is the arithmetic unit, while the bottom three represent the memory pipeline, which supports up to three memory operations in flight under certain constraints.
The colored regions represent which kind of instruction is at the top of the reorder buffer in the vector unit.
We observe that the register spilling takes half of the total execution time of the kernel.

\begin{figure}[!htbp]
    \centering
    \includegraphics[width=.9\columnwidth]{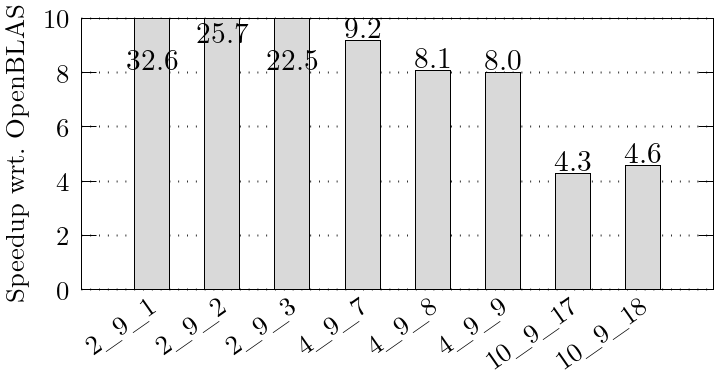}
    \caption{Speedup of batched GEMM with respect to non-batched OpenBLAS.}
    \label{figSeissolBenchmarkPerformance}
\end{figure}

\paragraph{Performance evaluation}
We wrote a synthetic benchmark suite that performs a number of GEMM calls of a set of matrix sizes given at compile time.
The benchmark compares the performance of our batched implementation with a loop of calls to the OpenBLAS library (\texttt{cblas\_dgemm}).
Figure~\ref{figSeissolBenchmarkPerformance} shows the speedup of our implementation with respect to the OpenBLAS reference.
Each bar represents a different matrix size and access type and we chose the use cases that are relevant for the SeisSol-proxy application.
We observe that our implementation beats the reference with speedups that range between~$4.3\times$ and~$32.6\times$.
The best cases correspond to small matrix sizes which have less register spilling.
A part from the spilling, another limiting factor of our implementation are the vector memory instructions \texttt{vlxe.v} and \texttt{vlse.v}.

\begin{figure}[!htbp]
    \centering
    \includegraphics[width=\columnwidth]{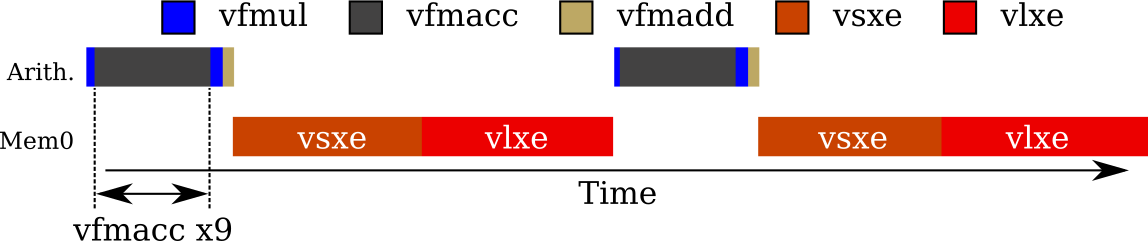}
    \caption{Instruction timeline of \texttt{20\_9\_10\_csi} (useful computation).}
    \label{figSeissolTimelineCompute}
\end{figure}

Figure~\ref{figSeissolTimelineCompute} shows a timeline of the case \texttt{20\_9\_10\_csi} zooming into the useful computation part.
In this region, we observe an alternating pattern of arithmetic instructions and memory instructions.
The timeline only includes one row for memory because the hardware does not overlap more than one indexed memory operation.
Thus, apart from register spilling, the use of \texttt{vlxe.v} and \texttt{vlse.v} is the other main limiting factor of our implementation.
Although we minimize the amount of such instructions, our measurements show that each one costs up to~$850$ cycles.
With the current layout of data structures in SeisSol, we cannot circumvent the use of indexed memory operations.
Future work includes redefining the data layout.

\subsection{Integration with SeisSol}\label{secIntegration}

\paragraph{Data structures}

\begin{figure}[!htbp]
    \centering
    \includegraphics[width=\columnwidth]{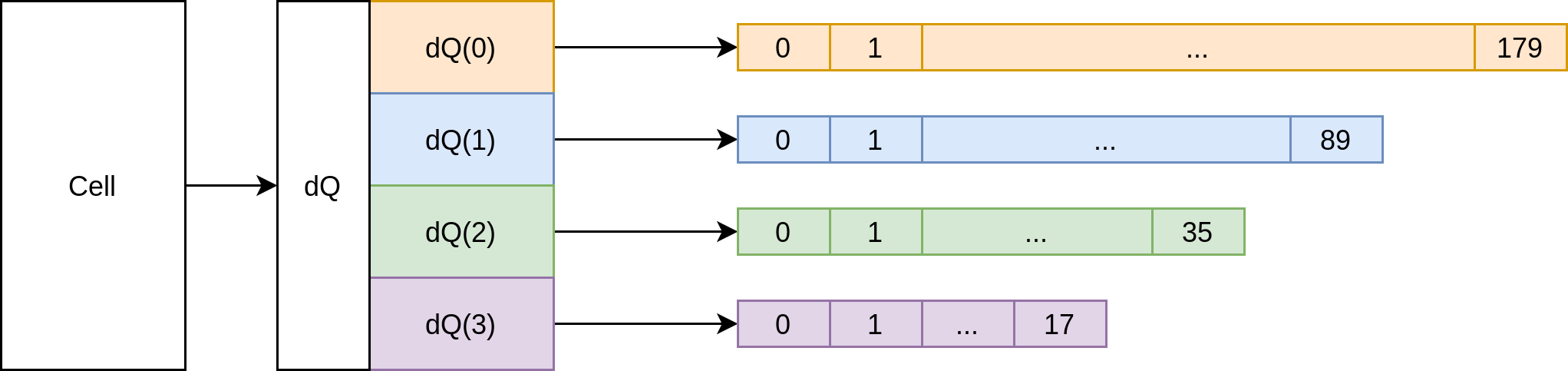}
    \caption{Example diagram of a tensor data structure in SeisSol (dQ).}
    \label{figSeissolDq}
\end{figure}

The relevant data structures for our work are representations of the mathematical object known as tensor.
In SeisSol, all tensor objects are declared and implemented using code generated during the build process.
Each type of tensor varies in size, but the general structure remains the same for all.
Figure~\ref{figSeissolDq} shows an example of such data structures, \texttt{dQ}.
The tensor object is a fixed-size array of pointers to matrices.
The size of each matrix is also fixed and known at compile time.
Tensors are the inputs and outputs of the GEMM-based kernels.

Each cell in the physical system modeled by SeisSol has a set of attributes mapped to tensor objects.
However, tensor objects of the same type (\eg \texttt{dQ}) are not stored contiguous in memory.
This is the reason for which our GEMM implementation cannot leverage strided memory accesses.

\paragraph{Code changes}

\begin{figure}[!htbp]
    \centering
    \includegraphics[width=\columnwidth]{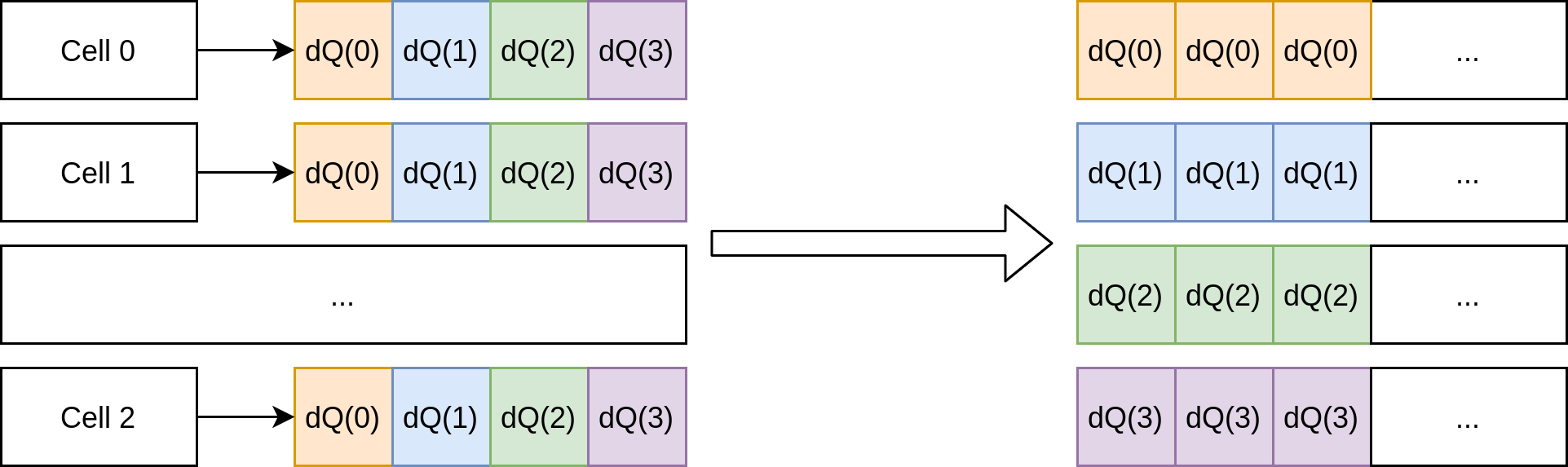}
    \caption{Data layout transformation to feed the batched GEMM function.}
    \label{figSeissolLayout}
\end{figure}

Firstly, we define and implement a batched version of each GEMM-based kernel.
For example, \texttt{execute} becomes \texttt{execute\_batched}.
Instead of taking two tensor objects (\texttt{kDivMT} and \texttt{star}) as inputs, and one tensor (\texttt{dQ}) as output,
our implementation takes in an array of pointers to matrices, in a data layout that is friendlier to our batched GEMM implementation.
Figure~\ref{figSeissolLayout} shows a schematic view of the layout transformation that we perform.

Secondly, we implement a new function called {\em computeLocalIntegrationBatched} which mirrors the already existing function {\em computeLocalIntegration}.
The body of the original function has two nested loops:
the top one iterates over all cells in the system,
while the innermost iterates through the components of tensor objects (\eg $[0,3]$ in the case of \texttt{dQ}).
Each iteration of the innermost loop calls the GEMM-based kernels.

Our batched version of the function performs three code modifications:
swap the order of the loops so that the one iterating through the cells becomes the inner-most;
transform the layout of the tensor objects to accommodate the batched kernels;
and call the batched versions of the GEMM-based kernels.

Thirdly, we add a memory allocation step before each timestep.
This phase allocates a buffer big enough to hold temporary data during the execution of the GEMM-based kernels.
In the reference implementation, this buffer was stack-allocated with a fixed size;
but the batched version requires it to be dynamically allocated because its size depends on the number of cells that are simulated, which is a runtime parameter.

Lastly, we add a runtime parameter option to the SeisSol-proxy app to choose which version should run:
scalar (reference), or
vector (batched).

\paragraph{Validation}

The SeisSol-proxy app does not perform any kind of validation.
We implement our own validation by writing to a file the contents of the tensor object \texttt{dQ} of each cell and compare the output between the reference and the batched versions.
In our tests, the difference between pairs of double precision elements was always under~$10^{-6}$, so we conclude that our code modifications output the same results as the reference.

\paragraph{Performance evaluation}

\begin{figure}[!htbp]
    \centering
    \includegraphics[width=.9\columnwidth]{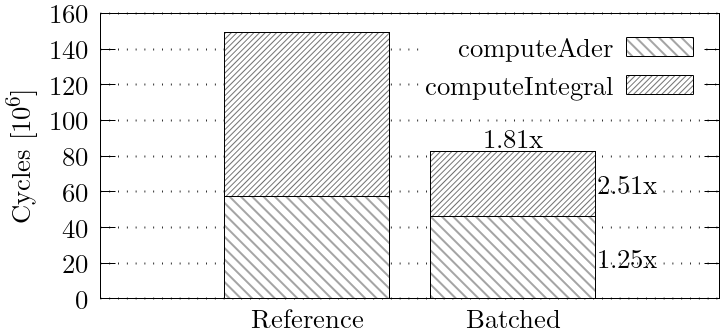}
    \caption{Cycles comparison between reference and batched versions.}
    \label{figSeissolIntegrationCycles}
\end{figure}

Figure~\ref{figSeissolIntegrationCycles} shows a performance comparison of the \texttt{computeLocalIntegration} function between the reference version using OpenBLAS and the batched version.
We observe a total speedup of~$1.81\times$ which is mainly achieved by the \texttt{computeIntegral} function.
This function calls the GEMM-based kernels with the biggest matrices.
For this reason, it is the most time consuming function and also the one in which our implementation suffers from spilling the most.

We know that the GEMM-based kernels represent~$53.53\%$ of the cycles in the \texttt{computeLocalIntegration} function.
By applying Amdahl's law, we calculate that the maximum overall speedup of SeisSol that can be achieved by only optimizing these kernels is~$2.15\times$ so we still have some room for improvement in our implementation.
However, the reader should note that all our code modifications and implementation of the batched GEMMs are written in plain C without any micro-architecture specific code.

\subsection{Porting to other architectures}\label{secPorting}
In this section we present the performance results of the same benchmark shown in Section~\ref{secBatchedDgemm}, Figure~\ref{figSeissolBenchmarkPerformance}, but running on \mn.
This is the flagship supercomputer at the Barcelona Supercomputing Center.
It is based on the Intel Sapphire Rapids CPU and supports~AVX-512 instructions.

\begin{figure}[!htbp]
    \centering
    \includegraphics[width=.9\columnwidth]{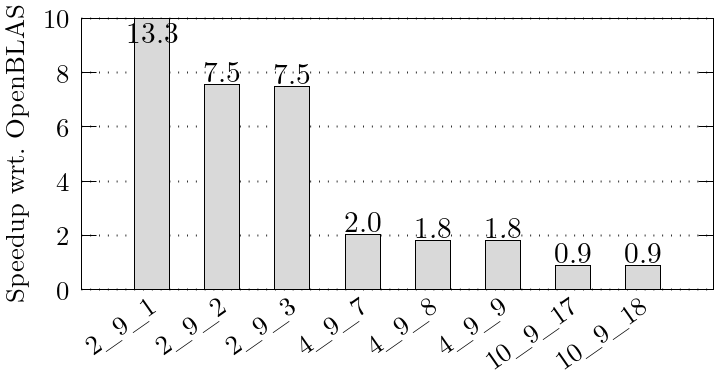}
    \caption{Speedup of batched GEMM with respect to OpenBLAS in \mn.}
    \label{figSeissolPerformanceMare5}
\end{figure}

Figure~\ref{figSeissolPerformanceMare5} shows the speedup of our batched GEMM implementation with respect to the OpenBLAS library specifically compiled for the core micro-architecture.
We observe a similar trend as with EPAC:
the bigger the matrices, the lower the speedup.
This is again caused by the register spilling, which is even more noticeable in the x86 architecture since it has less registers available than RISC-V.
With input sizes \texttt{10\_9\_17} and \texttt{10\_9\_18}, our batched version yields~$10\%$ less performance compared to the reference.
Nonetheless, we are able to achieve better performance with small matrices and having made no code modifications to our library.
The code is fully portable between CPU architectures.

\section{MiniFALL3D}\label{secMiniFall3D}

\subsection{Initial Study}

\fall~\cite{folch_2022_6343786} is an Eulerian model for the atmospheric transport and ground deposition of passive particles.
Although the model was originally developed for volcanic particles such as tephra, it has been extended to include mineral dust, aerosols ($SO_{2}$, $H_{2}O$) and radionuclides.

\fall is a multi-purpose model, as it can be used to compute both the airborne concentration (\eg at flight levels) and the fallout deposit (ground accumulation).
It is also a multi-scale model: it can run from local scales of few kilometers to continental scales of thousands of kilometers.
For each particle class, \fall solves the so-called Advection-Diffusion-Sedimentation (ADS) equation.
It also implements both 1st order Eluler and 4th order Runge-Kutta time-stepping methods.

In this work, we perform our experiments with \mfall \cite{mf3d-repo}, a CPU-only mini-app based on the main \fall 8.2.1 code.
Both app and mini-app are written in modern, object-oriented Fortran-90.%

\paragraph{Build configuration} For building \mfall, we disabled MPI parallelization by using the \texttt{--disable-parallel} configure flag.
We also used the \texttt{NETCDF} environment variable to point to our NetCDF-Fortran installation.
In \mfall, this library is only used for writing the output results of the simulation and has no impact on the performance measurements.

Execution parameters can be configured at runtime via a \texttt{params.inp} input file.
We run the default problem of type \texttt{UNIFORM}, which contains a single particle class and a default grid size of $100\times100\times60$.

The length of the simulation can be configured with the \texttt{RUN\_END\_(HOURS)} setting.
We also added a custom \texttt{SDV\_IITERS} switch to achieve finer grained control and directly indicate the number of timesteps to be run.

We set the \texttt{LOG\_LEVEL} settings to disable NetCDF output, since it is not required in our timing runs.

\begin{figure}[htbp]
  \centering
  \includegraphics[width=\linewidth]{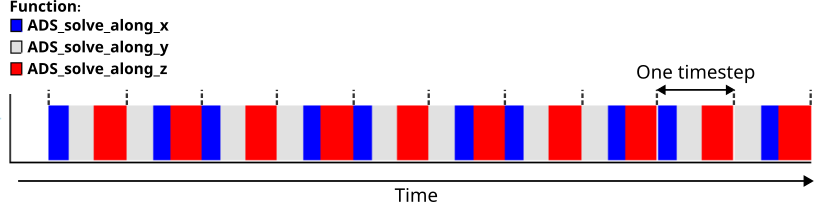}
\caption{Timeline of 10 timesteps in \mfall, highlighting ADS solver functions.}
  \label{fig10Timesteps}
\end{figure}

\paragraph{Execution structure} The execution of \mfall is structured into timesteps.
Each timestep contains a call to each one of the three ADS solver functions: \texttt{ADS\allowbreak\_solve\allowbreak\_along\allowbreak\_x}, \texttt{ADS\_solve\_along\_y} and \texttt{ADS\_solve\_along\_z}.
We notice that the first function in a timestep alternates between \texttt{ADS\allowbreak\_solve\allowbreak\_along\allowbreak\_x} and \texttt{ADS\_solve\_along\_y}.
Figure~\ref{fig10Timesteps} shows a timeline of a \mfall scalar execution limited to 10 timesteps.
The $x$-axis represents time, while colored regions correspond to the execution of the ADS solver functions.

There is a small white initialization interval previous to the first timestep, which can span up to 5\% of our 10 timestep shortened run.
For this reason, we chose to exclude initialization time from our reports, since it would be negligible in a full length 145-timestep execution (simulation spanning 24 hours).

\begin{figure}[htbp]
  \centering
  \includegraphics[width=\linewidth]{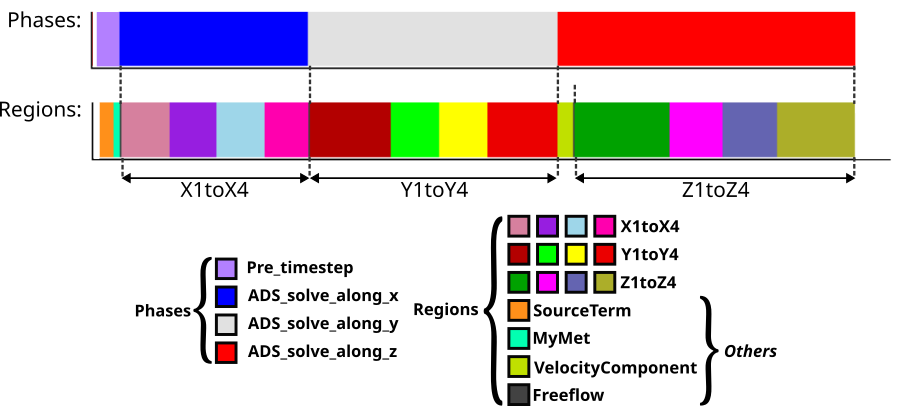}
  \caption{Timeline of one timestep in \mfall. Phases appear on top, while regions of study appear in the bottom.}
  \label{fig1Timesteps}
\end{figure}

The timeline at the top of Figure~\ref{fig1Timesteps} shows a close view of one timestep of a \mfall scalar execution.
We can identify four different phases.
We have one phase for each of the three known ADS solver functions, with the addition of a {\em Pre\_Timestep} phase.
This phase includes the computations performed at the beginning of a timestep, which fall outside of the scope of the ADS solver functions.
The bottom timeline of Figure~\ref{fig1Timesteps}, we provide a finer-grained decomposition of the timestep into a set of 16 regions.
In this study, we group these 16 regions into four distinct region groups.

The first three region groups are {\em X1toX4}, {\em Y1toY4} and {\em Z1toZ4}.
Since the ADS solver functions implement a 4th order Runge-Kutta method (RK4), each group includes four regions, one for each RK4 stage.
The last group is named {\em Other}, and contains all other regions which fall outside the RK4 solver.
It includes regions {\em SourceTerm} and {\em MyMet}, from the {\em Pre\_timestep} phase;
{\em VelocityComponent}, at the beginning of {\em ADS\_solve\_along\_z}; and {\em Freeflow}, a very short region executed once inside all ADS solver functions.

\paragraph{Performance out of the box} Here we report the performance of the vanilla \mfall mini-app running 10~timesteps.
We produced two builds: a scalar build and an autovectorized build, labeled \textit{0.vanillaSca} and \textit{0.vanilla}.
Execution time is reported as time spent inside our defined regions, in order to exclude initialization time.

The scalar execution of \mfall took ~$22\times10^6$ cycles,
while the auto-vectorized execution took~$17\times10^6$ cycles.
This corresponds to a speedup of $1.27\times$ with auto-vectorization with respect to scalar execution.
Since this speedup is unusually low,  we examine the potential issues that may be limiting the performance of the auto-vectorized code, in order to try to overcome them.

\paragraph{Factors limiting Vector Mix and Vector Activity}
Vector Mix tells us about the portion of executed code which is being vectorized, while Vector Activity tells us about the portion of time we spend running vector instructions.

For Vector Mix, we see that only 0.36\% of the executed instructions use vectors.
As a consequence, Vector Activity shows us we are using vector instructions only 9.05\% of the time.
Based on experiments on a variety of codes and benchmarks, we generally observe that, in order to achieve significant performance improvements, Vector Mix should be higher than 20\%, while Vector Activity should be higher than 80\%.

\paragraph{Factors limiting VL} Here we examine the Average Vector Length (AVL) of vector instructions found inside our regions.
Even though there are enough elements in our $100\times100\times60$ problem grid to fill a vector register,
out study reveals that, for X1toX4 and Y1toY4, the AVL is~106, and for Z1toZ4 is~68.
This shows we are underusing our VPU, since it can hold~256 double-precision elements in a vector register and process them with a single vector instruction.

Listing~\ref{figX1pseudocode} shows a pseudocode excerpt of the first phase of the RK4 solver found in the X1toX4 region, which we call {\em X1}.

\begin{lstlisting}[language=Fortran, caption=Excerpt of X1 (RK4 phase 1 from X1toX4), label=figX1pseudocode]
; RK4 phase 1 nested loops
do k=my_kps,my_kpe ! 1,60
  do j=my_jps,my_jpe ! 1,100
    call KT_RHS
    do i=my_ips_2h,my_ipe_2h; ... ; end do ! -1,102
    do i=my_ips,my_ipe; ... ; end do ! 1,100
  end do
end do


; KT_RHS function
function KT_RHS
  do i=ips-1,ipe+1; ... ; end do ! 0,101
  do i=ips-1,ipe+1; ... ; end do ! 0,101
  do i=ips,ipe ! 1,100
    ! ...
    if (i.eq.ips) then ; ... ; end if
    if (i.eq.ipe) then ; ... ; end if
  end do
\end{lstlisting}

We observe that the basic structure of the X1 region is two nested loops $k$ and $j$ which iterate over the $Z$ and $Y$ dimensions of the input.
Then, on the body of the nested loops, we have many 1-dimensional $i$ loops which iterate over the $X$ dimension, some of which are actually encased by the \texttt{KT\_RHS} function.

In this code excerpt, only the innermost $i$ loops are auto-vectorized.
Looking at their loop boundaries, we see that these loops iterate over 100, 102 or 104 elements.
This is the reason why an AVL of 106 is being reported in the execution.

Inside the Y1toY4 region, we observe the same limitation, but with $j$ as the innermost variable.
In Z1toZ4 the innermost variable is $k$, which iterates over the $Z$ dimension, we see an AVL close to 68.
Therefore, we conclude that in \mfall, the VL is limited by the size of the input in each dimension.

A possible solution to overcome this one-dimensional limitation is to
collapse the inner $i$ loops with the middle $j$ loops.
The collapsed loop would iterate over $100\times104$ elements, exposing enough parallelism to fill vector registers up to 256 elements.
However, the MPI implementation of the RK4 method incorporates special halo elements which is exchanged across MPI processes that share neighboring regions of the domain decomposition during the transition from one RK4 stage to the next one.

The disparity in loop boundaries (1:100, 0:101, -1:102) limits our ability to implement the loop fusion.
Combining loop collapsing with the extra logic to account for the halo introduces masked vector instructions, which are very costly in terms of cycles.
Furthermore, the mask would need to be different at each iteration.
In short, overcoming the 1-dimensional VL limitation would require rewriting part of the ADS code and its associated data structures.
These changes could help expose more of the parallelism intrinsic to the problem,
but they require extensive rewrite of application code.
Therefore, in this work we do not improve the AVL on the ADS solver and we instead focus on other factors limiting the performance.

\paragraph{Use of long latency instructions} Sometimes, an extensive use of costly vector operations such as reductions or slides can be a limiting factor.
This is not the case of \mfall, in which the most costly operation type is the \texttt{vfdiv} vector floating point division.
After examining the code, we found that vector division instructions are being placed efficiently by the compiler and cannot be avoided.

\subsection{Improving Vector Mix in the solver}
\label{secSolver}

Previously, we saw that the low Vector Mix values could be behind the small performance improvements obtained with the \textit{0.vanilla} auto-vectorized version.
In this section, we try to find which parts of the execution are responsible for the low Vector Mix and whether if it is possible to improve it by modifying the source code.
Table~\ref{tabX1toX4} summarizes all the relevant performance metrics of each code version that are discussed from hereon.
Versions are cumulative and build up on the changes of their predecessors.
For clarity, version names include a number indicating their order.

\begin{table}[htbp]
\centering
\caption{Metrics for region X1toX4 with different code versions.} \label{tabX1toX4}
\resizebox{\columnwidth}{!}{%
  \begin{tabular}{r|rrrrr}
  \multicolumn{1}{c|}{X1toX4} & \begin{turn}{45}0.vanillaSca\end{turn} & \begin{turn}{45}0.vanilla\end{turn} & \begin{turn}{45}1.vecKtrhs\end{turn} & \begin{turn}{45}2.lowlevel\end{turn} & \begin{turn}{45}3.subr\end{turn} \\ \hline
  cycles $[10^{9}]$ & 5.49 & 3.93 & 2.09 & 1.40 & 0.92 \\
  speedup & 1.00 & 1.40 & 2.63 & 3.93 & 5.94 \\
  Vmix & 0.00\% & 0.44\% & 1.81\% & 3.27\% & 7.94\% \\
  Vact & 0.0\% & 11.0\% & 31.4\% & 47.9\% & 72.4\% \\
  scaInst $[10^{9}]$ & 5.21 & 3.33 & 1.34 & 0.77 & 0.32 \\
  vecInst $[10^{6}]$ & 0.00 & 14.64 & 24.24 & 25.20 & 25.20
  \end{tabular}%
}
\end{table}

For simplicity, Table~\ref{tabX1toX4} only shows metrics for the X1toX4 region, since our three defined regions (X1toX4, Y1toY4, Z1toZ4) show a high degree of similarity.

\begin{figure}[htbp]
  \centering
  \includegraphics[width=\linewidth]{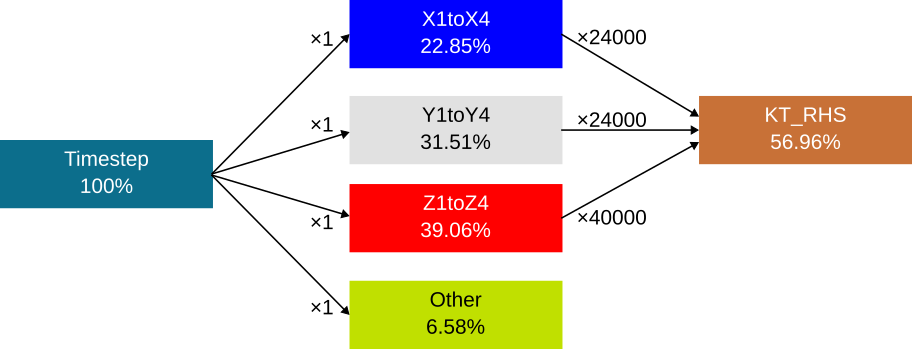}
  \caption{Number of calls and weights (\% of total cycles) in \textit{0.vanilla}.}
  \label{figWeightKT}
\end{figure}

\paragraph{1.vecKtrhs: Increasing vectorization in the KT\_RHS function}
As seen in Listing~\ref{figX1pseudocode}, the \texttt{KT\_RHS} function contains conditional clauses inside the innermost loop.
The compiler is not able to auto-vectorize this Fortran high-level construct for our target hardware.
This limitation has a high impact on overall performance, since the \texttt{KT\_RHS} function is called inside the loop body of all stages of the RK4 solver.
Figure~\ref{figWeightKT} shows a graph indicating the number of function calls to \texttt{KT\_RHS} of each region.

Even though a single call to \texttt{KT\_RHS} is very short, in aggregate it accumulates~56.96\% of total execution cycles.
The graph also shows that \texttt{KT\_RHS} is called thousands of times from all three X1toX4, Y1toY4 and Z1toZ4 regions.
This high call count comes from the fact that \texttt{KT\_RHS} is called inside the inner loop body of the RK4 solver. 

From our study of the vanilla version of the mini-app, we know it presents Vector Mix issues which leads to low performance.
And moreover, we know that \texttt{KT\_RHS} consumes a high percentage of time of the execution.
Therefore, if we manage to increase the Vector Mix of such a time consuming function, we de-facto contribute to improve the Vector Mix of the whole mini-app and achieve better performance.

\begin{lstlisting}[language=Fortran, caption=0.vanilla, label=figLstKTRHS]
do i = ips,ipe
   ! Compute F_p, F_m, P_p, P_m
   KT_RHS(i) = -1.0_rp*(F_p-F_m)/dxb(i)  + (P_p-P_m)/dxb(i)
   if(i.eq.ips) flux(1)=F_m+P_m
   if(i.eq.ipe) flux(2)=F_p+P_p
end do
\end{lstlisting}

Listing~\ref{figLstKTRHS} shows the original Fortran code of the \texttt{KT\_RHS}.
We observe that the conditional clauses depend on the loop control variable \texttt{i} and are only triggered on the first and last iterations.
In a scalar or SIMD processor, this \texttt{if} construct inside the loop is natural and could have some benefits, such as being able to re-use costly computations from the loop body like the \texttt{F\_m}, \texttt{P\_m}, \texttt{F\_p}, \texttt{P\_p} local variables.
But in a vector processor, it is difficult to reproduce this approach efficiently.
As a rule of thumb, it is good practice to avoid auto-vectorized loops that write scalar variables without following a clear reduction pattern.
In this particular case, the compiler is unable to generate vector code for our hardware.

In order to avoid the conditionals inside the loop, we opt for extracting them and have one non-conditional statement before the loop and another one after the loop.
The only drawback of this approach is that we cannot reuse computations from the loop body, and we need to expand local variables into their corresponding expression.
Listing~\ref{figLstKTRHSv2} shows this new implementation of the \texttt{KT\_RHS} function, which is refered to as \textit{1.vecKtrhs} in Table~\ref{tabX1toX4} and from hereon.

\begin{lstlisting}[language=Fortran, caption=1.vecKtrhs, label=figLstKTRHSv2]
flux(1) = 0.5_rp*(c_l(ips)* ! ...
do i = ips,ipe
   ! Compute F_p, F_m, P_p, P_m
   KT_RHS(i) = -1.0_rp*(F_p-F_m)/dxb(i)  + (P_p-P_m)/dxb(i)
end do
flux(2) = 0.5_rp*(c_l(ipe+1)*+ ! ...
\end{lstlisting}

Columns \textit{0.vanilla} and \textit{1.vecKtrhs} from Table~\ref{tabX1toX4}, show that the speedup (compared to \textit{0.vanillaSca}) improves from~$1.40\times$ to~$2.63\times$ after applying the changes described in this section.
This increase in performance can be attributed to the improved vectorization of previously scalar code in the \texttt{KT\_RHS} function, since we see both a decrease in scalar instructions and an increase in vector instructions.

Even after the reported improvements, Vector Mix and Vector Activity are still far from our target values of 20\% and 80\%, although all loops in our main regions are vectorized.
Therefore, we keep analyzing the solver for other possible sources of inefficiency.

\paragraph{2.lowlevel: Cutting off scalar instructions in loops}

In version \textit{1.vecKtrhs}, all loops of the main regions (Z1toX4, Y1toY4 and Z1toZ4) are vectorized.
However, a closer inspection of execution traces reveals that there are scalar instructions being executed in between loops.

\begin{figure}[htbp]
  \includegraphics[width=\linewidth]{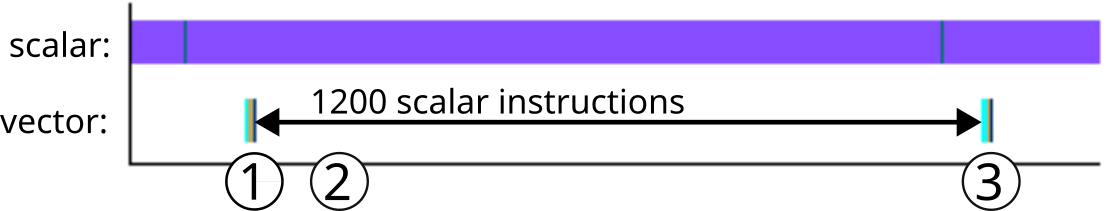}
  \begin{lstlisting}[language=Fortran]
! 1) first loop
c0(my_ips:my_ipe,j,k) = my_c(my_ips:my_ipe,j,k,ibin)
    + 0.5_rp*dt*ki(my_ips:my_ipe)
! 2) although there are no statements,
!    we find 1200 scalar instructions here
! 3) second loop
new_c(my_ips:my_ipe,j,k) = new_c(my_ips:my_ipe,j,k)
    + dt*ki(my_ips:my_ipe) / 3.0_rp\end{lstlisting}
  \caption{X2 subregion: showing scalar instructions between vector loops.}
  \label{figLowlevelrave}
\end{figure}

Figure~\ref{figLowlevelrave} shows a zoomed timeline of two loops from the X1toX4 region with the corresponding excerpt of source code.
The upper part of the timeline is colored in purple when a scalar instructions is executed, while the bottom part is colored when a vector instruction is executed.
There are only two colored segments of vector instructions, which are marked as (1) and (3) in the timeline.
These correspond to the two loops seen in the code excerpt.

The space between the two loops is labeled as (2) in the timeline. Even though we see in the code that there are no statements between the loops, we observe 1200 scalar instructions in place.
Where do these scalar instructions come from?

An important aspect of the code is that the loops are written as Fortran array operations, in which the whole array can be operated in a single statement.
The input and output arrays might overlap in memory position (commonly referred to as {\em aliasing}) so the compiler takes a conservative approach to ensure correctness and generates code to account for aliasing.
One such approach is to copy the input array into a temporary location, removing the possibility for aliasing.
We suspect that the scalar instructions we are seeing in Figure~\ref{figLowlevelrave} correspond to scalar copies of this kind triggered by the compiler.

From the programmers perspective, we know that there is no aliasing between the arrays.
In our case, excess scalar instructions originate from the usage of high-level Fortran array constructs.
When these are replaced with equivalent low-level \texttt{do-loops}, the excess scalar instructions disappear.
Array traversal direction is explicit in Fortran do-loops, so the compiler does not need to worry about possible overlapping.

\begin{lstlisting}[language=Fortran, caption=Replacing high level array operations with equivalent do-loops., label=figLstLowlevel]
! 1.vecKtrhs
c0(my_ips:my_ipe,j,k) = my_c(my_ips:my_ipe,j,k,ibin)
    + 0.5_rp*dt*ki(my_ips:my_ipe)

! 2.lowlevel
do epi_i = my_ips,my_ipe
   c0(epi_i,j,k) = my_c(epi_i,j,k,ibin) + 0.5_rp*dt*ki(epi_i)
end do
\end{lstlisting}

We replaced all array operations found inside our Z1toX4, Y1toY4 and Z1toZ4 regions, as seen in the example from Listing~\ref{figLstLowlevel}.
In the \textit{1.vecKtrhs} and \textit{2.lowlevel} columns from Table~\ref{tabX1toX4}, we see that speedup compared to the vanilla scalar version increases significantly from~$2.63\times$ to~$3.93\times$.
Almost half of the executed scalar instructions are removed (from 133 million to 77 million), while vector instructions have only increased slightly (from 24 to 25 million).
The reduction of scalar instructions can be attributed to the use of do-loops, which do not trigger scalar copies.
We also observed a small increase in vector instructions due to a slight variation on the choice of instructions by the compiler, although it is not significant.
We keep analyzing the solver, as we still find that values of Vector Mix and Vector Activity are still far from the target values of 20\% and 80\%.

\paragraph{3.subr: Replacing functions with subroutines}
We already detected excess scalar instructions surrounding high-level array operations.
Our investigations also highlighted a similar occurrence regarding function calls.

In \mfall, functions are used thoroughly inside the solver in order to encapsulate and reuse common code across several ADS stages.
Previously, we acknowledged the importance of the \texttt{KT\_RHS} function, called in the loop body of all ADS stages.
Its first loop is actually enclosed inside another \texttt{r\_sbee} function, although it was omitted from Listing~\ref{figX1pseudocode} for simplicity.
In Figure~\ref{figWeightKT}, we saw that inside the X1toX4 region, \texttt{KT\_RHS} (and consequently, \texttt{r\_sbee}) are called tens of thousands times.
Therefore, after considering such high call count, we see the scalar instructions which surround them can have significant impact in the overall execution.

In Fortran, there are two types of procedures: functions and subroutines.
Functions allow a return argument, while subroutines do not.
This scheme is very flexible, since the return argument can be both input/output and also an array.

In \mfall, functions \texttt{KT\_RHS} and \texttt{r\_sbee} have an array input/output argument.
We know from the previous \textit{2.lowlevel} version that high level array constructs can be problematic, as the compiler may need to account for memory aliasing.
We modified the code by transforming these functions into equivalent subroutines.
After changing the keyword, the return argument also needs to be changed into a regular input/output argument.
This modification is very simple and only involves changing in the procedure header and calls, but not its body.

\begin{lstlisting}[language=Fortran, caption=Replacing functions with subroutines. \textit{KT\_RHS} return becomes argument \textit{retu}., label=figLstSubr]
! 2.lowlevel
function KT_RHS(c,dxp,dxb,u,k,flux,ips,ipe,ibs,ibe)
  ! ...
  real(rp), intent(inout) :: flux(2)
  real(rp) :: KT_RHS(ips:ipe)

! 3.subr
subroutine KT_RHS_subr(c,dxp,dxb,u,k,flux,ips,ipe,ibs,ibe,retu)
  ! ...
  real(rp), intent(inout) :: flux(2)
  real(rp), intent(inout) :: retu(ips:ipe)
\end{lstlisting}

Listing~\ref{figLstSubr} shows the \texttt{KT\_RHS} original function header and the new subroutine version labeled \textit{3.subr}. Some function arguments have been omitted for clarity.
In the upper part we see the header original function, with the \textit{KT\_RHS} return argument.
Return arguments must have the same name as the function, and do not apear in the argument list.
On the lower part appears the equivalent subroutine, in which the return argument has been converted to the regular argument \textit{retu}.
Although not shown here, we performed this change on function \texttt{r\_sbee} as well.

Looking at columns \textit{2.lowlevel} and \textit{3.subr} on Table~\ref{tabX1toX4} we see that speedup compared to vanilla scalar increases from~$3.93\times$ to~$5.94\times$.
Now, more than half the executed scalar instructions have been removed (from 77 to 31 million), while vector instructions remain completely unchanged.
This reduction of scalar instruction has affected Vector Mix and Vector Activity, which stand at 7.94\% and 72.4\%, closer to the minimum target value of 80\%.

\subsection{Improving Vector Mix in the Other regions}
\label{secOther}

\begin{figure}[htbp]
\centering
  \includegraphics[width=0.8\linewidth]{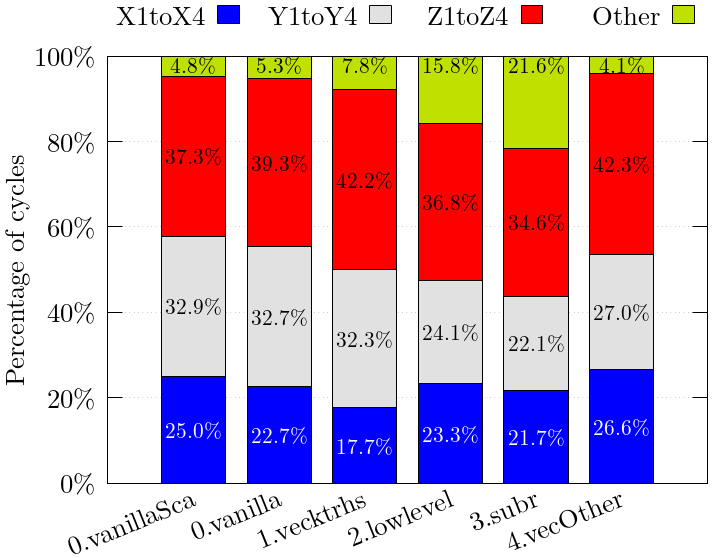}
  \caption{Cycle distribution across region groups in \mfall}
  \label{figPlotCycDistrib}
\end{figure}

When defining our regions of study on Figure~\ref{fig1Timesteps}, we defined a set of four subregions grouped under the \textit{Other} label.
Figure~\ref{figPlotCycDistrib} shows the cycle distribution between region groups accross different code versions.

We can see how in the \textit{0.vanilla} vectorized version, the \textit{Other} regions span only 5.3\% of the total execution cycles in a timestep.
But now, after applying the code changes described in Section~\ref{secSolver}, we have improved Vector Mix in the solver regions.
Therefore, their amount of cycles has grown smaller, while for the unmodified \textit{Other}, cycles have remained unchanged.
We also see that, for version \textit{3.subr}, \textit{Other} takes 21.6\% of timestep execution cycles.
This region presents a Vector Mix and Vector Activity of 0.2\% and 9.2\%, so its vectorization has an enormous potential for improvement, which is explored in this section.

\paragraph{Flattening multidimensional arrays to increase VL}
The compiler used in this study only supports vectorizing the innermost loop, even with perfectly nested loops.
This is a limiting factor, since ideally Vector Length should depend on the total work performed on the whole set of nested loops, not just on the innermost loop.
This pattern can be found inside the two loops that form the \textit{MyMet} region, and in one of the two loops in \textit{VelocityComponent}, which we show in Listing~\ref{figFlattening}.

\begin{lstlisting}[language=Fortran, caption=3-D array operation and equivalent 1-D do-loop., label=figFlattening]
! 3D array construct
CB%


! Equivalent 1-D loop implemented with do-loops
! Computing 1D array size
ijksize = (my_kbe_1h-my_kbs_1h+1) * (my_ipe-my_ips+1)
    * (my_jpe-my_jps+1)
! Pointer assignments
cb_w_1d(1:ijksize) => CB%
my_w1_1d(1:ijksize) => my_w1(:,:,:)
my_w2_1d(1:ijksize) => my_w2(:,:,:)
! 1D do-loop
do i=1,ijksize
  cb_w_1d(i) = (1.0_rp-stime) * my_w1_1d(i) + stime * my_w2_1d(i)
end do
\end{lstlisting}

All these loops iterate over 3-D arrays in a compact manner using Fortran array operations.
We can use Fortran pointers to define a pointer associated to the 3-D array, which allows traversing it as if it were 1-D.

In this particular case, the loop iterates over the three dimensions of the input matrix.
After our transformation, Vector Length is only limited by the whole size of the input, not just one dimension.
In our $100\times100\times60$, this allows reaching the maximum Vector Length available in our hardware of 256 double-precision elements.

\paragraph{Loop reordering to increase VL}
Listing~\ref{figLoopSwap} shows a simplified view of the loop found in the \textit{SourceTerm} region, which is also limited to the VL of the innermost loop.

\begin{lstlisting}[language=Fortran, caption=Swapping loops \texttt{ibin} and \texttt{i} in \textit{SourceTerm}., label=figLoopSwap]
! Original loop ordering: i -> ibin
do k = my_kps,my_kpe
  ! compute dZ
  do j = my_jps,my_jpe
    ! compute dY, Hm1, Hm2
    do i = my_ips,my_ipe ! 1,100
       ! compute dX, Hm3
       do ibin = 1,MY_TRA%
          vol = dX*dY*dZ*(Hm1*Hm2*Hm3)
          my_c(i,j,k,ibin) = ! ...

! Loops swapped: ibin -> i
do k = my_kps,my_kpe
  ! compute dZ
  do j = my_jps,my_jpe
    ! compute dY, Hm1, Hm2
    do ibin = 1,MY_TRA%
      do i = my_ips,my_ipe ! 1,100
        ! compute dX, Hm3
        vol = dX*dY*dZ*(Hm1*Hm2*Hm3)
        my_c(i,j,k,ibin) = ! ...
\end{lstlisting}

In this case, the innermost induction variable \texttt{ibin}, iterates over the particle classes defined in the input.
This scheme enables the re-use of local variables between the \texttt{ibin} iterations,
but in our case is only one.
To overcome this limitation, we chose to change the ordering of the loops.
We swapped the \texttt{ibin} and \texttt{i} loops, so now the innermost loop iterates over the $X$ dimension of the input, which limits our VL to 100 instead of the 1 from \texttt{ibin}.
In this case, it is not possible to further increase the VL with a flattening transformation, since the loops are not perfectly nested.

\paragraph{Vectorizing previously scalar code}
The \textit{freeflow} region contains multiple 1-D loops, such as the one seen in Listing~\ref{figOmpSimd}, which are not being vectorized by the compiler.
We used the \texttt{!\$omp simd} directive to enforce auto-vectorization of these loops, which resulted in a fully vectorized \textit{freeflow} region.
We also observed that in this region, tuples were represented as arrays of size two, which are read inside the loop bodies.
The compiler failed to detect tuples as loop-invariant, so we simply assigned their value to individual local variables, which results in a more efficient code for the loop body.

\begin{lstlisting}[language=Fortran, caption=\textit{freeflow}: Adding directive and changing tuple into variable., label=figOmpSimd]
! Original code: unable to auto-vectorize
do j = my_jps,my_jpe
   ! ...
   c(my_ips-1,j,k) = 2.0_rp*CB%
end do

! Modified code: successfully auto-vectorized
bvalue1 = CB%
!$omp simd
do j = my_jps,my_jpe
   ! ...
   c(my_ips-1,j,k) = 2.0_rp*bvalue1-c(my_ips,j,k)
end do
\end{lstlisting}

All modifications presented in this section (flattening multidimensional arrays, loop reordering, and vectorizing the {\em freeflow} region) are grouped together into the final code version of \mfall presented in this work: {\em 4.vecOther}.

The following section summarizes the performance benefits of each code version running on EPAC and other systems which are based on different architectures.

\subsection{Performance evaluation}

All plots in this section aim to visualize the performance improvements that each code version provides in three different systems.
The stacked columns show the amount of cycles spent in each region (left $y$-axis).
The line shows the speedup with respect to the vanilla code with auto-vectorization enabled (right $y$-axis, first point is always~1).
The reader should note that the left $y$-axis (cycles) changes scales depending on the system,
but we keep the scale of the right $y$-axis (speedup) constant for all plots to highlight the benefits of our code modifications with respect to the vanilla version.

\begin{figure}[htbp]
  \centering
  \includegraphics[width=0.8\linewidth]{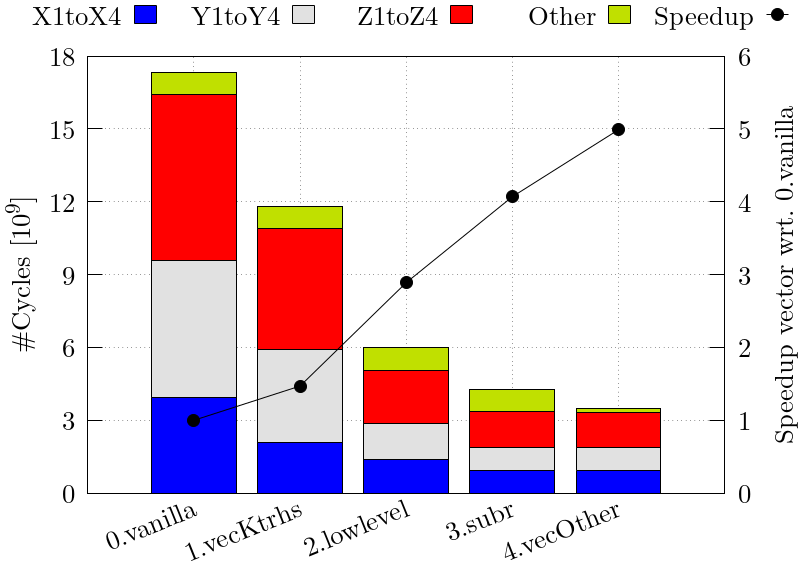}
  \caption{Cycles and speedup of \mfall in EPAC}
  \label{figPlotSDVSameCode} 
\end{figure}

\paragraph{Performance on EPAC} Figure~\ref{figPlotSDVSameCode} shows the performance of \mfall when running on EPAC.
Looking at the execution cycles in the stacked plot, we see that the most impactful code modifications are \textit{2.lowlevel} and \textit{3.subr}, since they remove a lot of scalar instructions.
We can also observe how the cycles of the \textit{Other} region remain constant until the last version, since no prior code modifications affected this region. 
In summary, after all our code modifications, version \textit{4.vecOther} is able to achieve a speedup of~$5\times$ with respect to \textit{0.vanilla}.

\begin{figure}[htbp]
  \centering
  \includegraphics[width=0.8\linewidth]{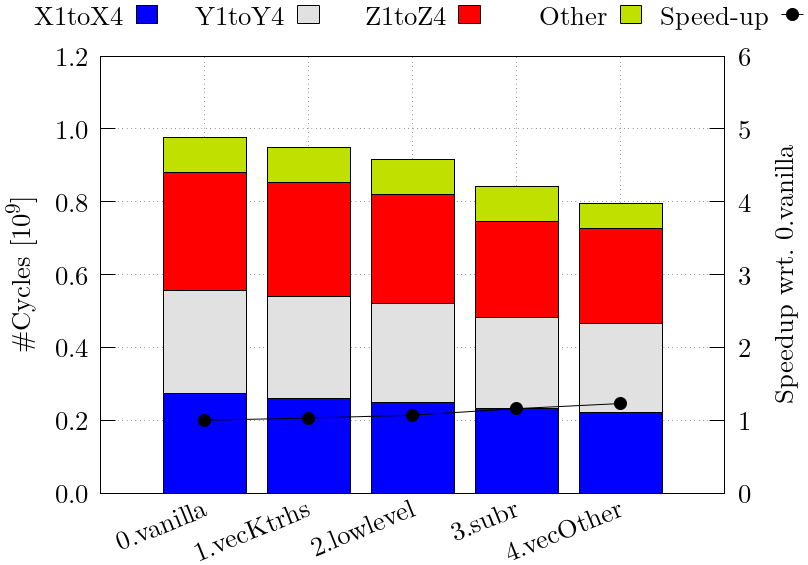}
  \caption{Cycles and speedup of \mfall in \mn}
  \label{figPlotMN5} 
\end{figure}

\paragraph{Performance on \mn} Figure~\ref{figPlotMN5} shows the performance of \mfall when running on \mn.
The CPU of this system supports AVX512, which is able to process up to eight double-precision elements with a single instruction.

The plot shows that all the code changes we developed targeting a long vector architecture do not compromise the performance of a SIMD machine like \mn.
Moreover, all versions contribute in some degree to the decrease of execution cycles.
Even though, the total speedup of~$1.23\times$ achieved with the best version is much lower than its EPAC counterpart of~$5\times$.

\begin{figure}[htbp]
  \centering
  \includegraphics[width=0.8\linewidth]{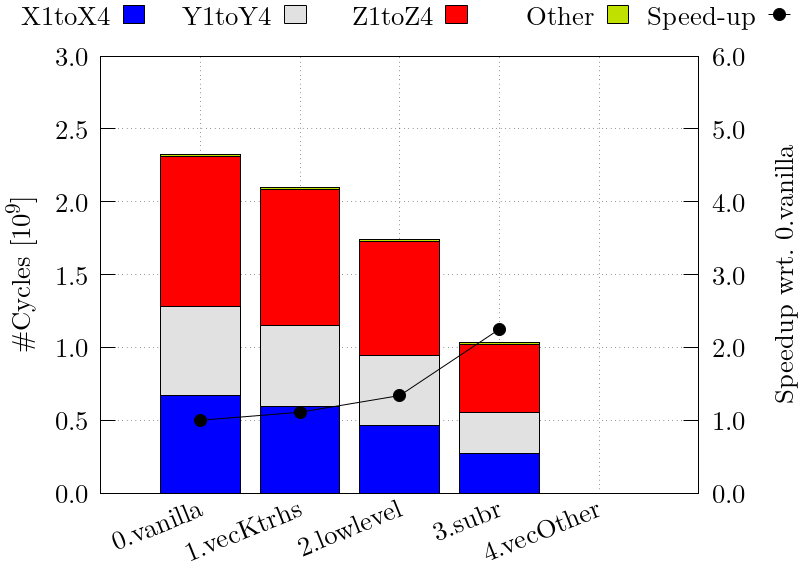}
  \caption{Cycles and speedup of \mfall in NEC}
  \label{figPlotNEC} 
\end{figure}

\paragraph{Performance on NEC} Figure~\ref{figPlotNEC} shows the performance of \mfall in the NEC SX Aurora Tsubasa long vector accelerator.
Its compiler also has auto-vectorization capabilities.

We were not able to compile the \textit{4.vecOther} version with the NEC \texttt{nfort} Fortran compiler.
In particular, the compiler rejected the pointer assignments that allow accessing a 3-D array as if it was 1-D.
Regardless of the compiler limitations, we observe that the weight in cycles of the \textit{Others} phase is very small compared to the rest, so there is not much of a global performance gain to be achieved by \textit{4.vecOther}.

As in EPAC, the two code versions which show the biggest leaps in performance are \textit{2.lowlevel} and \textit{3.subr}, and the speedup achieved with the best version stands at~$2.25\times$.

\section{Related Work}\label{secRelatedWork}

The use of RISC-V vector extensions in high-performance computing (HPC) is now well-established in scientific literature.
Diehl et al. in~\cite{Diehl2024Preparing} investigate astrophysics codes,
Blancafort et al. in~\cite{blancafort2024exploiting} study a fluid dynamics code,
and Torres et al. in~\cite{Torres2024Codesigning} examine materials science codes.

At a lower level of software development, there are also numerous efforts to vectorize libraries for RISC-V, demonstrating a growing interest in RISC-V and vectorization not only at a level of scientific application.
Rani Gupta et al. in~\cite{Gupta2023Challenges} present a study on the vectorization of convolution algorithms on RISC-V vector architectures.
Vizcaino et al. in~\cite{Vizcaino2022Acceleration} explore the acceleration of FFT kernels, while in~\cite{Vizcaino2024Graph}, they introduce a library for the vectorization of graph-related problems.

The choice made in our work to rely on compiler autovectorization appears to be the right path to follow, as analyzed by Adit et al. in~\cite{Adit2022Performance}. In their study, they emphasize that there is still room for performance gains to be exploited within compilers supporting variable vector length architectures. This suggests that having scientific codes prepared for autovectorization may enable further performance improvements as compilers become more advanced.

Regarding the FPGA-based platform used for the evaluation, the EPI project advocates for fast prototyping on FPGAs. This approach also seems to be adopted by other research groups.
Islam et al., in~\cite{Islam2023Resource}, use an FPGA implementation of a vector RISC-V core to perform convolutional neural network operations with 8-bit datatypes. Our approach complements that of~\cite{Islam2023Resource}, as our design is more oriented toward high-performance computing (HPC).

Finally, concerning the literature related to the selected codes in the context of Earth sciences,
there have been previous optimization efforts on SeisSol,
particularly in batching kernel executions targeting GPUs~\cite{10.1145/3432261.3436753}.
Our work diverges from these earlier efforts by implementing batched kernels targeting CPUs and ensuring the code remains portable across different architectures.
Regarding \mfall, the novelty of this work lies in the performance study of \mfall on a long-vector platform.
Previous experiments have focused on optimizing the code in~\cite{Folch2019FALL3D} and adding GPU acceleration in a different version of the mini-app using OpenACC directives~\cite{mf3d-talk}.
Our work complements these earlier studies by focusing on acceleration using vector architectures and, once again, demonstrating performance portability when compared to other HPC architectures.

\section{Conclusions}\label{secConclusions}

In this study, we successfully ported two earth science codes, SeisSol and \mfall, to EPAC, a RISC-V-based system with an integrated vector unit that can process~256 double precision elements per instruction.
Through the application of the evaluation methodology outlined in~\cite{short-reasons}, we were able to identify key areas for improvement in both codes and apply techniques to improve their performance without hindering the portability of the code.
All the code modifications proposed in this work are high-level (plain C and Fortran) which yield better performance in EPAC and NEC.
At worst, our modifications do not introduce any performance penalties in other systems such as \mn.

In the case of SeisSol, our results show that exposing more work to the compiler can lead to significant performance gains.
This is a case of how improving the average Vector Length of the vectorized code better leverages hardware based on long-verctors.
We proposed and implemented a batched GEMM library targeting small matrices achieving a maximum speedup of $32.6\times$ with respect to the reference.
However, this approach also resulted in increased memory footprint and register spilling.
Future work includes applying a heuristic to change between implementation depending on matrix sizes so that bigger matrices do not cause register spilling but still may benefit from long-vector architectures.

In \mfall, we demonstrated the effectiveness of incremental code changes in improving performance, with a speedup of~$5\times$ achieved through targeted optimizations.
This is a case of how to identify and improve low Vector Mix and Vector Activity in a complex scientific code.
We also identified the importance of considering high-level Fortran constructs, such as array operations and functions, which may limit compiler optimizations.

The findings of this study contribute to our understanding of the challenges and opportunities in porting earth science codes to emerging architectures like RISC-V.
By demonstrating the potential for significant performance gains through optimized code modifications, we hope to inspire further research into the development of high-performance computing solutions for these applications.

\section{Acknowledgments}\label{secAcks}

Supported by the EuroHPC Joint Undertaking (JU): FPA N. 800928 (EPI), SGA N. 101036168 (EPI-SGA2), and GA N. 101093038 (ChEESE-2P CoE). The JU receives support from the EU Horizon 2020 research and innovation programme and from Croatia, France, Germany, Greece, Italy, Netherlands, Portugal, Spain, Sweden, Denmark and Switzerland.  The EPI-SGA2 project, PCI2022-132935 is also co-funded by MCIN/AEI /10.13039/501100011033 and by the UE NextGeneration EU/PRTR.
Supported by the pre-doctoral program AGAUR-FI ajuts (2024 FI-200424) Joan Oró offered by Secretaria d'Universitats i Recerca del Departament de Recerca i Universitats de la Generalitat de Catalunya.
Special thanks for their kind support on SeisSol internals to Sebastian Wolf and Michael Bader from the Technical University of Munich - School of Computation, Information and Technology.

\bibliography{999-bibliography}

\end{document}